\documentclass[twocolumn,prb,amsmath,amssymb,floatfix]{revtex4}

\usepackage{graphicx}
\usepackage{dcolumn}
\usepackage{bm}

\begin{document}


\title{Amplitude-Phase Coupling in a Spin-Torque Nano-Oscillator}

\author{Kiwamu Kudo}
\email{kiwamu.kudo@toshiba.co.jp}
\author{Tazumi Nagasawa}
\author{Rie Sato}
\author{Koichi Mizushima}
\affiliation{Corporate Research and Development Center, Toshiba Corporation, Kawasaki, 212-8582, Japan}

\date{\today}

\begin{abstract}
The spin-torque nano-oscillator in the presence of thermal fluctuation 
is described by the normal form of the Hopf bifurcation with an additive white noise. 
By the application of the reduction method, 
the amplitude-phase coupling factor, 
which has a significant effect on the power spectrum of the spin-torque nano-oscillator, 
is calculated from the Landau-Lifshitz-Gilbert-Slonczewski equation 
with the nonlinear Gilbert damping. 
The amplitude-phase coupling factor exhibits a large variation 
depending on in-plane anisotropy under the practical external fields.
\end{abstract}


\maketitle

When a direct current $I$ flows into a magnetoresistive (MR) device, 
a stationary magnetic state becomes unstable and 
a steady magnetic oscillation is excited by the spin-transfer torque. 
The oscillation is expected to be applicable to a nanoscale microwave source, i.e., 
the spin-torque nano-oscillator (STNO).\cite{Kiselev,Rippard} 
According to the theory 
based on the spin-wave Hamiltonian formalism,\cite{Slavin,Tiberkevich,Kim,Kim2} 
the frequency nonlinearity plays a key role in determining the behavior of the oscillator. 
It has been shown that 
the strong frequency nonlinearity leads to 
significant effects on the power spectrum of STNO 
in the presence of thermal fluctuation: 
a linewidth enhancement\cite{Kim} 
and non-Lorentzian lineshapes\cite{Kim2}. 
In this paper, the important nonlinearity is examined. 
From the Landau-Lifshitz-Gilbert-Slonczewski (LLGS) equation as the model of STNO, 
we calculate explicitly the magnitude of the quantity 
corresponding to the normalized frequency nonlinearity $N\slash \Gamma_\mathrm{eff}$ 
(see, e.g., Eq.~(4) in Ref.~\onlinecite{Kim2}) of the spin-wave approach. 
In particular, we take account of in-plane anisotropy of a magnetic film which has 
been neglected in the early studies\cite{Slavin,Tiberkevich,Kim,Kim2}, 
finding the large effect of the anisotropy on the nonlinearity.

We describe STNO by a generic oscillator model. 
It is known that 
small-amplitude oscillations near the Hopf bifurcation point are 
generally governed by the simple evolution equation 
for a complex variable $W(t)$ 
known as 
the Stuart-Landau (SL) equation.\cite{Kuramoto} 
The SL equation is 
derived as a normal form of the supercritical Hopf bifurcation 
from the general system of ordinary differential equations. 
Accordingly, the LLGS equation similarly reduces to the SL equation 
in the case where the Hopf bifurcation, 
which represents a generation of magnetic oscillations in STNO, occurs. 
The reduction of the LLGS equation can be executed by the reduction method 
based on the center-manifold theorem. 
At finite temperature, 
there exists inevitable thermal magnetization fluctuation in STNO.\cite{Kim3,Mizushima} 
We include the thermal effect into the magnetization dynamics 
by just adding white noise term to the SL equation, i.e., 
STNO in the presence of thermal fluctuation is described 
by the `noisy' Hopf normal form: 
\begin{equation}
\frac{\mathrm{d} \tilde{W} }{\mathrm{d} \tilde{t} }
=i \tilde{\Omega} \tilde{W}+(1+i\delta )(p-|\tilde{W}|^2) \tilde{W} 
+\eta(\tilde{t}),
\label{eq:NHN2}
\end{equation}
where $\tilde{W}$ is the normalized complex variable representing 
the amplitude and phase of a magnetization vector $\bm{M}$ 
(see Eq.~(\ref{eq:neutral}) below). 
In Eq.~(\ref{eq:NHN2}), 
$\tilde{\Omega}$ represents a fundamental frequency, 
$\tilde{t}$ is a normalized dimensionless time, and 
$\eta(\tilde{t})$ is the zero-mean, white Gaussian noise 
with the only non-vanishing second moment given by 
$\langle \eta(\tilde{t}) \bar{\eta}(\tilde{t}') \rangle
=4 \delta(\tilde{t}-\tilde{t}')$. 
$p$ is the bifurcation parameter. 
An oscillation is generated when $p$ becomes positive. 
In the context of STNO, $p \propto (I-I_\mathrm{c})$ where 
$I_\mathrm{c}$ is the threshold current. 
The parameter $\delta$ quantifies the coupling between the amplitude and phase fluctuations and 
is called the {\it amplitude-phase coupling factor}. 
It is $\delta$ that we calculate numerically in this paper 
and that corresponds to the normalized frequency nonlinearity $N\slash \Gamma_\mathrm{eff}$ 
of the spin-wave approach. 
The amplitude-phase coupling factor $\delta$ affects the power spectrum of an oscillator 
and leads to linewidth enhancement and non-Lorentzian lineshapes.\cite{Risken,Gleeson} 
Due to its effect, the factor $\delta$ is also called 
the {\it linewidth enhancement factor}.\cite{Henry} 
Eq.~(\ref{eq:NHN2}) is often used as the simplest model of a noisy 
auto-oscillator 
in many fields, for example, electrical engineering, chemical reactions, optics, biology, 
and so on.\cite{Risken,Haken} 
Therefore, we can easily compare STNO with conventional oscillators and 
clarify its features.

The amplitude-phase coupling factor $\delta$ is obtained 
in the procedure of the reduction of the LLGS equation. 
In the following, we first explain the LLGS equation. 
Then, following Kuramoto's monograph\cite{Kuramoto}, 
we consider an instability of a steady solution and 
execute the reduction of the LLGS equation.

\begin{figure}
\includegraphics[width=4cm]{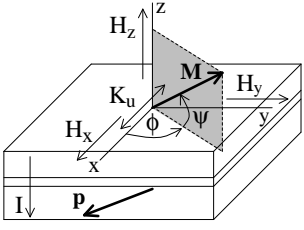}
\caption{The spherical coordinate system $(\phi,\psi)$ for 
the direction of the free layer magnetization $\bm{m}=\bm{M}\slash M_\mathrm{s}$ of STNO. 
$\bm{p}$ denotes the direction of the pinned layer magnetization; 
$\bm{p}=(\cos \psi_\mathrm{p} \cos \phi_\mathrm{p}, 
\cos \psi_\mathrm{p} \sin \phi_\mathrm{p}, \sin \psi_\mathrm{p})$. }
\label{fig:spherical}
\end{figure}

The magnetic energy density of the free layer of STNO is assumed to 
have the form 
\begin{equation}
\mathcal{E}=
-\bm{M} \cdot \bm{H}_\mathrm{ext}
-\frac{K_\mathrm{u}}{M_\mathrm{s}^2} (\bm{M} \cdot \hat{\bm{x}})^2
+\frac{1}{2} 4 \pi \bm{M} \cdot \mathcal{N} \cdot \bm{M}, 
\label{eq:MagEnergy}
\end{equation}
where $M_\mathrm{s}$ is the saturation magnetization, 
$\bm{H}_\mathrm{ext}=H_x\hat{\bm{x}}+H_y\hat{\bm{y}}+H_z\hat{\bm{z}}$ is an external field, 
$K_\mathrm{u}$ is uniaxial anisotropy along the $x$ direction, 
and $\mathcal{N}$ is the demagnetizing tensor; 
$\mathcal{N}=\mathrm{diag}(N_x,N_y,N_z)$. 
Using the spherical coordinate system (see Fig.~\ref{fig:spherical}), 
we describe the magnetization dynamics of STNO by the LLGS equation 
\begin{equation}
\begin{cases}
\cos \psi~ \dot{\phi}
=-\alpha(\xi) \dot{\psi} 
-F_1(\phi,\psi,\omega_J)
\\
\dot{\psi}
=\alpha(\xi) \cos \psi~\dot{\phi} 
+F_2(\phi,\psi,\omega_J)
\end{cases},
\label{eq:LLGsp}
\end{equation}
where 
$F_1(\phi,\psi,\omega_J) \equiv 
(\gamma\slash M_\mathrm{s}) \partial \mathcal{E}\slash \partial \psi -a(\phi,\omega_J)$ and 
$F_2(\phi,\psi,\omega_J) \equiv 
(\gamma \slash (M_\mathrm{s} \cos \psi)) \partial \mathcal{E}\slash \partial \phi
+b(\phi,\psi,\omega_J)$. $\gamma$ is the gyromagnetic ratio. 
The second terms of $F_i$ result from the Slonczewski term
$\bm{T}_J=(\gamma a_J\slash M_\mathrm{s})\bm{M} \times (\bm{M} \times \bm{p})$ 
in which $a_J$ is proportional to the current density $J$ 
through the free layer\cite{Slonczewski}. 
Therefore, 
$a(\phi,\omega_J)\equiv \omega_J \cos \psi_\mathrm{p} \sin(\phi-\phi_\mathrm{p})$ 
and 
$b(\phi,\psi,\omega_J)
\equiv \omega_J [ \cos \psi_\mathrm{p} \sin \psi \cos (\phi -\phi_\mathrm{p}) 
- \sin \psi_\mathrm{p} \cos \psi ]$, where $\omega_J =\gamma a_J$. 
$\alpha(\xi)$-terms of Eqs.~(\ref{eq:LLGsp}) are the generalized Gilbert damping terms 
proposed by Tiberkevich and Slavin.\cite{Tiberkevich2} 
We take into account only the first non-trivial term of the 
Taylor series expansion for $\alpha(\xi)$ by the magnetization change rate 
$\xi\equiv (\partial \bm{m}\slash \partial t)^2 \slash (\gamma 4 \pi M_\mathrm{s})^2$; 
$\alpha(\xi)=\alpha_\mathrm{G}(1+q_1 \xi)$. 
According to Ref.~\onlinecite{Tiberkevich2}, the nonlinear LLGS model with $q_1=3$ 
gives a good agreement with the experimental results 
of Ref.~\onlinecite{Kiselev} and Ref.~\onlinecite{Mistral}.

An instability of a steady solution of Eq.~(\ref{eq:LLGsp}) is considered. 
A steady solution $(\phi_0(\omega_J),\psi_0(\omega_J))$ 
is derived from $F_i(\phi_0,\psi_0,\omega_J)=0$. 
Shifting the variables as $u_1 \equiv \phi-\phi_0$ and $u_2 \equiv \psi-\psi_0$, 
we have the Taylor series of Eq.~(\ref{eq:LLGsp}) as follows, 
\begin{equation}
\dot{\bm{u}}
=L \bm{u} + N_2 \bm{u} \bm{u}+N_3 \bm{u} \bm{u} \bm{u}+ \cdots
\label{eq:Taylor1}
\end{equation}
where $\bm{u}=(u_1,u_2)^\mathrm{T}$. 
Here, the diadic and triadic notations~\cite{Kuramoto} have been used. 
The stability of a steady solution is determined by the eigenvalues of 
the linear coefficient matrix $L$: $\lambda_\pm=\Gamma \pm (\Gamma^2 - \det L)^{1\slash 2}$. 
$\Gamma$ is defined as 
$\Gamma =\Gamma(\omega_J)\equiv (1\slash 2) \mathrm{tr} L$ 
and plays the role as a control parameter since it depends on $\omega_J$. 
We confine ourselves to the case where the Hopf bifurcation occurs. 
Then, $\lambda_\pm$ is a pair of complex-conjugate eigenvalues. 
The point, $\Gamma=0$, is the Hopf bifurcation point; 
while a steady solution remains stable for $\Gamma<0$, 
it becomes unstable for $\Gamma>0$. 
The bifurcation point corresponds to the threshold 
$\omega_J^\mathrm{c}$ which is determined by 
$\mathrm{tr}L=0$ and $F_i(\phi_0,\psi_0,\omega_J^\mathrm{c})=0$. 
Near the bifurcation point, we divide $L$ into 
the two parts; $L=L_0+\Gamma L_1$, 
where $L_0$ is the critical part and $\Gamma L_1$ is the remaining part. 
Corresponding to $L$, $\lambda_{+}$ is also divided into the two parts; 
$\lambda_{+}=\lambda_0+ \Gamma \lambda_1$. Although 
$L_1$ and $\lambda_1$ generally depend on $\Gamma$ further, 
we neglect their dependence and evaluate them by the values at $\Gamma=0$. 
Accordingly, $\lambda_0= i \omega_0$ 
and 
\begin{equation}
\lambda_1=1-\frac{1}{2i \omega_0} 
\left. \frac{\mathrm{d}}{\mathrm{d} \Gamma } \det L \right|_{\Gamma=0},
\label{eq:lambda1}
\end{equation}
where $\omega_0 \equiv \sqrt{\det L_0}$. 
The right and left eigenvector of $L_0$ 
corresponding to the eigenvalue $\lambda_0$ are denoted 
as $\bm{U}$ and $\bm{U}^\ast$, respectively. 
These are normalized as 
$\bm{U}^\ast \bm{U}=\bar{\bm{U}^\ast} \bar{\bm{U}}=1$ where 
$\bar{\bm{U}}$ means a complex conjugate of $\bm{U}$.

Let us apply the reduction method to Eq.~(\ref{eq:Taylor1}). 
The SL equation for a complex amplitude $W(t)$, 
\begin{equation}
\dot{W}=\Gamma \lambda_1 W-g|W|^2 W
\label{eq:SL}
\end{equation}
and the neutral solution for the magnetization dynamics, 
\begin{equation}
\begin{pmatrix}
\phi \\
\psi
\end{pmatrix}
=
\begin{pmatrix}
\phi_0 \\
\psi_0
\end{pmatrix}
+W(t) e^{i \omega_0 t} \bm{U}
+\bar{W}(t) e^{-i \omega_0 t} \bar{\bm{U}}
\label{eq:neutral}
\end{equation}
are obtained 
within the lowest order approximation.\cite{Kuramoto} 
Under the approximation, only the Taylor expansion coefficients 
up to the third order are needed. 
The complex constant $g$ in Eq.~(\ref{eq:SL}) is given by 
\begin{equation}
\begin{split}
g\equiv \nu_1+i \nu_2=
& -3( \bm{U}^\ast, N_3 \bar{\bm{U}} \bm{U} \bm{U} ) \\
& +4( \bm{U}^\ast, N_2 \bm{U} \bm{V}_0 )
+2( \bm{U}^\ast, N_2 \bar{\bm{U}} \bm{V}_{+} ),
\end{split}
\end{equation}
where $\bm{V}_0=L_0^{-1} N_2 \bm{U} \bar{\bm{U}}$ and 
$\bm{V}_{+}=(L_0 -2 i \omega_0)^{-1} N_2 \bm{U} \bm{U}$. 
The amplitude-phase coupling factor $\delta$ is obtained from the complex constant $g$ 
and is given by 
\begin{equation}
\delta=\nu_2\slash \nu_1.
\label{eq:delta}
\end{equation}
In this way, the factor $\delta$ for STNO can be calculated numerically 
from the parameters of the LLGS equation. 

\begin{figure}
\includegraphics[width=4.2cm]{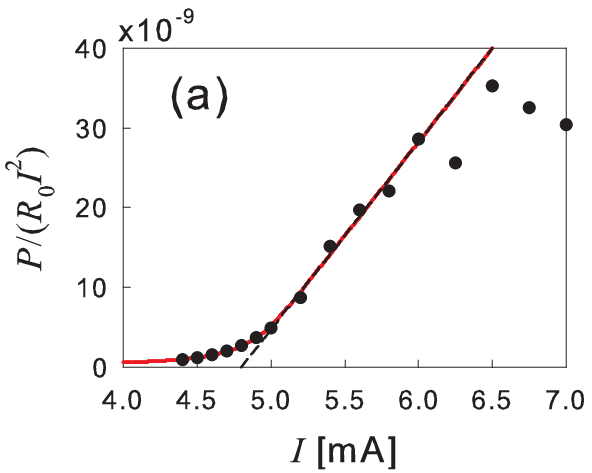}
\includegraphics[width=4.2cm]{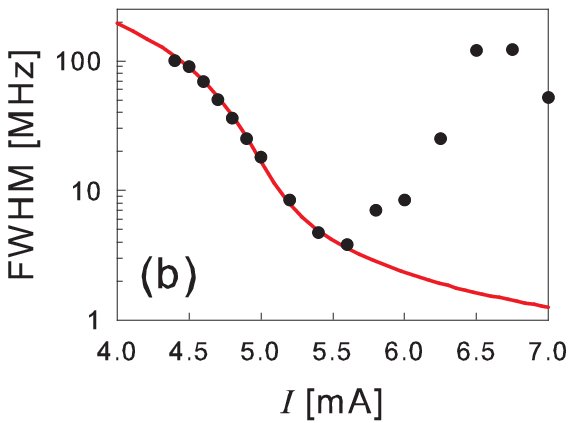}
\caption{\label{fig:compare} (Color online) (a) Power $P$ divided by $R_0I^2$ with 
$R_0=13.6~\Omega$ and (b) linewidth (FWHM) of the signal of STNO 
as a function of applied current $I$. 
Dots are experimental data at $T=150$~K taken from Ref.~\onlinecite{Mistral}. 
Red lines are theoretical fitting curves based on the model of Eq.~(\ref{eq:NHN2}). }
\end{figure}

The noisy Hopf normal form given by Eq.~(\ref{eq:NHN2}) is derived 
when we add the noise term $f(t)$ with $\langle f(t) \bar{f}(t') \rangle =4 D \gamma^2 \delta(t-t')$ 
to the SL Eq.~(\ref{eq:SL}). 
$f(t)$ has the dimension of frequency. 
The components in Eq.~(\ref{eq:NHN2}) are defined as 
$\tilde{W}(t)
=(D \gamma^2 \slash \nu_1)^{-1\slash 4} W(t) e^{i (\omega_0+\Gamma \delta-\Gamma \mathrm{Im}\lambda_1) t}$, 
$\tilde{t} \equiv \sqrt{D \gamma^2 \nu_1} t$, 
$p \equiv \Gamma \slash \sqrt{D \gamma^2 \nu_1}$, 
and $\tilde{\Omega}\equiv \omega_0\slash \sqrt{D \gamma^2 \nu_1}$. 
Therefore, we can make the most of many well-known properties of Eq.~(\ref{eq:NHN2})
\cite{Risken,Gleeson} 
to examine the behavior of STNO. 
It is known, for example, that 
the spectrum linewidth $\Delta \omega_\mathrm{FWHM}$ far above the threshold ($p \gg 0$) 
is increased by a factor of $(1+\delta^2)$.\cite{Risken} 
In the context of STNO, when $\Gamma \gg 0$, the linewidth 
can be expressed as 
\begin{equation}
\Delta \omega_\mathrm{FWHM}
=
\Delta \omega_\mathrm{res} 
\times \frac{k_\mathrm{B} T}{E_\mathrm{osci} } \times \frac{1}{2} (1+\delta^2),
\label{eq:FWHM2}
\end{equation}
which corresponds to Eq.~(11) in Ref.~\onlinecite{Kim}. 
Here, $k_\mathrm{B}T$ is the thermal energy. 
$\Delta \omega_\mathrm{res}$ is the linewidth at thermal equilibrium ($\omega_J=0$) given by 
$\Delta \omega_\mathrm{res}=2 \Gamma_\mathrm{eq}$, 
where $\Gamma_\mathrm{eq}\equiv -\Gamma(\omega_J=0)$. 
Moreover, $E_\mathrm{osci}$ is the magnetization oscillating energy and can be written as 
$E_\mathrm{osci}
\simeq 2 \bm{U}^\dagger [ \frac{ \partial( \partial_{u_1} \mathcal{E}, \partial_{u_2} \mathcal{E} )}
{ \partial (u_1,u_2) }]_{\bm{u}=\bm{0}} \bm{U} P_W V_\mathrm{free} 
=\frac{1}{2} \frac{\Gamma_\mathrm{eq} k_\mathrm{B} T}{D \gamma^2} P_W$ 
when it is assumed that $E_\mathrm{osci} \simeq k_\mathrm{B} T$ 
near thermal equilibrium ({\it energy equipartition}). 
Here, $V_\mathrm{free}$ is the volume of the free layer and 
$P_W$ is the total power of $W(t)$ given by 
$P_W=
\sqrt{D \gamma^2\slash \nu_1} 
\{ p+ 2\slash F(p)\}$ 
with $F(p)\equiv \sqrt{\pi} e^{p^2\slash 4} [1+\mathrm{erf}(p\slash 2)] $. 
From the expression of Eq.~(\ref{eq:FWHM2}), 
it is found that 
the MR device in STNO 
itself is nothing but a resonator on the analogy of 
electrical circuits. 
The other one of well-known properties of Eq.~(\ref{eq:NHN2}) is 
that the amplitude-phase coupling factor distorts the power spectrum 
to non-Lorentzian lineshapes especially near the threshold 
(see, e.g., FIG.~5 of Ref.~\onlinecite{Gleeson}). 
The degree of the lineshape distortion is determined by 
the magnitude of $\delta$ and $p$, 
corresponding to the calculation in Ref.~\onlinecite{Kim2}. 
We comment on the validity of Eq.~(\ref{eq:NHN2}) 
for large-amplitude oscillations. 
In Fig.~\ref{fig:compare}, the theoretical fitting curves based on the model Eq.~(\ref{eq:NHN2}) 
are compared with the experimental data of Ref.~\onlinecite{Mistral} 
and give a good agreement with them up to $I \simeq 5.6$ mA ($p\simeq 8.2$) beyond 
the threshold current $I_\mathrm{c}=4.8$ mA ($p=0$) estimated by the fitting.
\cite{note} 
Therefore, although the derivation of Eq.~(\ref{eq:NHN2}) 
is based on a perturbation expansion around the bifurcation point, 
it is considered to be valid for rather large-amplitude oscillations with $p\sim 10$.

We briefly mention the oscillating frequency $\omega_\mathrm{osci}$. 
From Eqs.~(\ref{eq:NHN2}) and (\ref{eq:neutral}), the oscillating frequency of 
a free layer magnetization far above threshold is written as 
$\omega_\mathrm{osci}=\omega_0-\Gamma \delta+ \Gamma \mathrm{Im} \lambda_1$. 
Although the calculation results for $\mathrm{Im} \lambda_1$ 
of Eq.~(\ref{eq:lambda1}) are not shown here, 
we have found that this quantity has a small value with $\mathrm{Im} \lambda_1 \sim \alpha_\mathrm{G}$ 
for wide range of parameters of the LLGS equation. 
Accordingly, $\omega_\mathrm{osci}$ is approximately given by 
$\omega_\mathrm{osci} \simeq \omega_0-\Gamma \delta$. 
Since $\Gamma \propto (I-I_\mathrm{c})$, 
while the frequency $\omega_\mathrm{osci}$ 
decreases as the current $I(>I_\mathrm{c})$ increases when $\delta>0$ (red shift), 
$\omega_\mathrm{osci}$ increases when $\delta<0$ (blue shift) 
in accordance with the spin-wave models~\cite{Slavin,Tiberkevich,Kim,Kim2}.

\begin{figure}
\includegraphics[width=8.3cm]{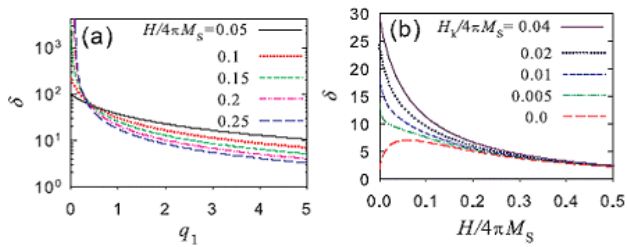}
\caption{\label{fig:in-plane}  (Color online) (a) Dependence of $\delta$ 
on the nonlinearity of the damping $q_1$ for various values of an external magnetic field $H$. 
An uniaxial anisotropy field is taken as $H_k\slash 4 \pi M_\mathrm{s}=0.04$. 
(b) Dependence of $\delta$ 
on an external magnetic field $H$ for various values of an uniaxial anisotropy field $H_k$.}
\end{figure}

As illustrated above, 
the amplitude-phase coupling factor $\delta$ plays a key role 
to determine the behavior of an oscillator. 
Therefore, the features of STNO can be found out by the calculation of $\delta$.

Some calculation examples of $\delta$ are shown in Fig.~\ref{fig:in-plane}. 
It is considered the case where a free layer is an in-plane magnetic film 
with an in-plane external field applied along the $x$ direction, $\bm{H}_\mathrm{ext}=H \hat{\bm{x}}$. 
It is assumed that $\mathcal{N}=\mathrm{diag}(0,0,1)$, $\alpha_\mathrm{G}=0.02$, 
and $(\phi_\mathrm{p},\psi_\mathrm{p})=(0,0)$. 
In Fig.~\ref{fig:in-plane}(a), the dependence of $\delta$ on the nonlinearity of the damping $q_1$ 
is shown. 
It is found that $\delta$ monotonically decreases for $q_1$ 
and the variation of $\delta$ is very large. 
This result suggests that a nonlinear damping significantly changes the LLG dynamics.
\cite{Tiberkevich2} 
In Fig.~\ref{fig:in-plane}(b), 
the dependence of $\delta$ on an external magnetic field $H$ 
for various values of an uniaxial anisotropy field $H_k(=2K_\mathrm{u}\slash M_\mathrm{s})$ is shown. 
The nonlinearity of the damping is taken as $q_1=3$.\cite{Tiberkevich2} 
In the practical external field region, 
$\delta$ is very sensitive to an uniaxial anisotropy field 
and varies largely. 
Therefore, when the dynamics of STNO is considered, 
it is necessary to take the effect of an uniaxial anisotropy field into account seriously. 
This is the main result of the present paper.

In summary, 
we have considered the dynamics of STNO by 
reducing the LLGS equation to a generic oscillator model 
and calculated explicitly the amplitude-phase coupling factor 
which is the key factor for the power spectrum. 
The amplitude-phase coupling factor $\delta$ is very sensitive to 
magnetic fields, in-plane anisotropy, and the nonlinearity of damping. 
The large variation of $\delta$ is the remarkable feature of STNO 
in comparison with conventional oscillators. 
The calculation way for $\delta$ shown 
is applicable for an arbitrary magnetization configuration 
and is useful for finding a stable STNO 
with small $\Delta \omega_\mathrm{FWHM}$ (Eq.~(\ref{eq:FWHM2})), 
which is preferable for applications.


\begin{thebibliography}{17}
\bibitem{note} 
The dimensionless power in Fig.~\ref{fig:compare}(a) is 
given by $P\slash R_0 I^2=a\{p+2\slash F(p)\}$ with $a\simeq 2.3063 \times 10^{-9}$ 
and $p\simeq 10.202(I-I_\mathrm{c})$. 
To obtain the linewidth in Fig.~\ref{fig:compare}(b), we have used the parameters of 
$\sqrt{D \gamma^2 \nu_1}\slash 2 \pi =11.24$ MHz and $\delta=0.5$, and have solved 
the eigenvalue problem of the Fokker-Planck equation corresponding to Eq.~(\ref{eq:NHN2}) 
as done in Ref.~\onlinecite{Risken} or Ref.~\onlinecite{Kim2}.
\end{thebibliography}
\end{document}